\documentclass[twocolumn,runningheads]{svjour2}
\smartqed  
\usepackage{graphicx}
\usepackage{graphics}
\journalname{Astrophysics and Space Science}
\begin{document}

\title{High-energy gamma-ray emission from the inner jet of LS~I+61 303: the hadronic contribution revisited
}


\author{M. Orellana\and G.~E. Romero}


\institute{M. Orellana \and
           G.~E. Romero \at
              Instituto Argentino de Radioastronom\'{\i}a\\ 
              C.C.5, (1894) Villa Elisa, Buenos Aires, Argentina\\
              \and Facultad de Ciencias Astron\'omicas y Geof\'{\i}sicas, 
              Universidad Nacional de La Plata, Paseo del Bosque s.n. (1900) La Plata, Argentina \\
              \email{morellana@carina.fcaglp.unlp.edu.ar}\\
              \email{romero@fcaglp.unlp.edu.ar}           
}

\date{Received: date / Accepted: date}

\maketitle

\begin{abstract}
LS~I+61 303 has been detected by the Cherenkov telescope MAGIC at very high energies, presenting a variable flux along the orbital motion with a maximum clearly separated from the periastron passage. In the light of the new observational constraints, we revisit the discussion of the production of high-energy gamma rays from particle interactions in the inner jet of this system. The hadronic contribution could represent a major fraction of the TeV emission detected from this source. The spectral energy distribution resulting from $pp$ interactions is recalculated. Opacity effects introduced by the photon fields of the primary star and the stellar decretion disk are shown to be essential in shaping the high-energy gamma-ray light curve at energies close to 200 GeV. We also present results of Monte Carlo simulations of the electromagnetic cascades developed very close to the periastron passage.
We conclude that a hadronic microquasar model for the gamma-ray emission in LS~I +61 303 can reproduce the main features of its observed high-energy $\gamma$-ray flux.
\end{abstract}

\section{Introduction}
\label{intro}
LS~I +61 303 is a Be X-ray binary with strong and variable radio emission. It has been suggested as the potential counterpart of the gamma-ray sources 2CG 135+01 and 3EG J0241+6103 (Gregory \& Talor 1978, Kniffen et al. 1997). Massi et al. (2001) detected relativistic jets in the source, which was then included in the microquasar class. The existence of jets was confirmed by the observations presented by Massi et al. (2004). Very recently, LS~I +61 303 was detected by MAGIC, a very large atmospheric imaging Cherenkov telescope located at La Palma (Albert et al. 2006). This makes LS~I +61 303 the second high-energy microquasar detected by ground-based telescopes. Moreover, MAGIC found clear evidence for variability in the detected signal: the source was stronger at orbital phases 0.5-0.6, whereas the periastron passage occurs at phase 0.23.

At X-rays the source has been detected by XMM-Newton, Beppo-SaX and INTEGRAL (Sidoli et al. 2006, Chernyakova et al. 2006) showing variability on short timescales and a harder spectrum when the source was brighter (Sidoli et al. 2006).

The orbital parameters of LS~I +61 303 have been determined by Casares et al. (2005). The eccentricity of the system is 0.72$\pm$0.15 and the orbital inclination is $i\sim30\pm20$ deg. The nature of the compact object is not well-established. A low-mass black hole cannot be ruled out (Massi 2004). If a mass of $\sim$ 2.5 M$_{\odot}$ is assumed for the compact object, then the mass of the Be star should be $\sim 12$ M$_{\odot}$, for $i=30$ deg. 

From a theoretical point of view, both accreting and non-accreting models have been proposed to explain the high-energy emission of LS~I+61 303. In the accreting scenario (e.g. Bosch-Ramon et al. 2006a) a relativistic jet is launched from the surroundings of the compact object. Strong magnetic fields are supported by an underluminous accretion disk, which is advection-dominated. High-energy emission can be produced in the jet by leptonic (e.g. Bosch-Ramon et al. 2006b, Gupta \& B\"ottcher 2006, Bednarek 2006a) or hadronic (Romero et al. 2003, 2005) interactions. In non-accreting models (Dubus 2006b, Chernyakova et al. 2006) particles are accelerated up to relativistic energies in the interacting region where a pulsar wind collides with the stellar wind. Inverse Compton (IC) cooling of electrons (Maraschi \& Treves 1981, Dubus 2006b) or $pp$ interactions (Chernyakova et al. 2006) could result in the production of high-energy photons.

A colliding wind model has been proposed as the mechanism producing the high-energy emission in the $\gamma$-ray binary PSR B1259-63, a system that contains a radio pulsar in an eccentric orbit around a Be star (Aharonian et al. 2005). In the case of LS~I+61 303, the existence of jets extending up to $\sim 400$ AU from the core seems to support an accretion/jet model (see the discussion in Mirabel 2006). Albert et al. (2006) favor a microquasar model and suggest that the detection of the high-energy signal after the periastron passage would support a leptonic origin for the emission. However, $\gamma$-ray propagation effects alone could be responsible for such a behavior (Bednarek 2006a).

In this paper we revisit the hadronic model for high-energy emission in LS~I+61 303 proposed by Romero et al. (2005) before the MAGIC detection. Using an improved accretion model and more sophisticated calculations for the opacity effects inside the binary system we show that a hadronic origin for the emission detected by MAGIC can not be ruled out.

\section{The model}
\label{sec:1} 
We consider gamma-ray production in hadronic interactions of relativistic protons from a jet launched close to the compact object and cold protons from the equatorial wind of the Be star (Romero et al. 2003, 2005).
The star has a slow equatorial wind that fits a density profile $\rho_{\rm w}(r)=\rho_0({r}/{R_*})^{-n}$, with $n=3.2$ and $\rho_0=5\times 10^{-11}$ g cm$^{-3}$ (Mart\'{\i} and Paredes 1995). The wind remains mainly near to the equatorial plane, confined in a disk with half-opening angle $\varphi=15^\circ$, and an initial outflowing radial velocity $v_{{\rm r}0}\sim 5$ km s$^{-1}$ (Mart\'{\i} and Paredes 1995). The disk effective temperature close to the star is $T_{\rm disk}=17000$ K. The values of the adopted parameters in our model are listed in Table \ref{tab}.

\begin{table*}[t]
  \caption[]{Model parameters}
  \label{tab}
  \centering
  \begin{tabular}{lllll}
  \hline\noalign{\smallskip}
 Parameter: description [units] &  values \\[3pt]
\tableheadseprule\hline\noalign{\smallskip}
$M_{\star}$: stellar mass [$M_{\odot}$] & 12 \\
$R_{\star}$: stellar radius [$R_{\odot}$] & 10  \\
$M_{\rm BH}$: compact object mass [$M_{\odot}$] & 2.5\\
$e$: eccentricity & 0.72 \\
$i$: orbital plane inclination [$^\circ$] & 30\\
$\omega$: angle of periastron [$^\circ$] & 20 \\
$R_{\rm 0}$: initial jet radius [$R_{\rm Sch}$] & 5 \\
$z_0$: jet initial point in the compact object RF [$R_{\rm Sch}$] & 50 \\
$\chi$: jet semi-opening angle tangent & 0.1 \\
$B_{\rm eq}$: equipartition magnetic field at the base of the jet [G] & $10^{8}$ \\
$\alpha$: relativistic proton power-law index & 2.5 \\
$\gamma_p^{\rm min}$: lowest Lorentz factor of relativistic protons & 2 \\
$\gamma_p^{\rm max}$: highest Lorentz factor of relativistic protons & $\sim 10^7$\\
$\Gamma$: macroscopic jet Lorentz factor & 1.25\\
$T_{\star}$: stellar surface temperature [K] & $26000$ \\
$T_{\rm disk}$: circumstellar disk temperature [K] & 17000 \\
$R_{\rm disk}$: circumstellar disk outer radius [$R_{\star}$] & 12 \\
$R_{\rm in}$: radius of the circumstellar disk brightest region [$R_{\star}$] & 3 \\
$L_{\rm disk}$: circumstellar disk luminosity [erg s$^{-1}$]& $2\times 10^{37}$\\
$\varphi$: disk half-opening angle [$^\circ$]& 15 \\
$v_{\rm r0}$: radial velocity of the wind at the base [cm s$^{-1}$] & $3\times 10^5$\\
$v_{\varphi0}$: azimuthal velocity of the wind at the base [cm s$^{-1}$] & $1.13\times 10^7/\sin i$\\
$\rho_0$: density of wind at the base [g cm$^{-3}$]& $5 \times10^{-11}$\\

  \noalign{\smallskip}\hline
  \end{tabular}
\end{table*}

The $pp$ interactions can occur either because there is some mixing of the stellar wind with the jet matter or because some protons escape from the jet into the wind. The level of interaction is quantified through a ``mixing factor", here assumed as $f_{\rm m}\propto v_{\rm rel}^{-1}\sim 0.1$. This phenomenological prescription accounts for a more efficient rejection of particles when a larger relative velocity is thought to enhance the boundary effects between the jet and the wind\footnote{The problem of
matter exchange through the boundary layers of a relativistic jet is a
difficult one. Its proper treatment is far beyond the scope of this work.}. At phase $\phi=0.5$ the mixing factor is maximum, reaching $f_{\rm m}\sim 0.5$.

The jet is assumed to be in a steady state. We are not dealing here with the details of the launching mechanism, which is supposed to be related to magneto-centrifugal effects (Blandford \& Payne 1982). The hadronic jet power is a fraction of the accretion power: $L_{\rm jet}=q_{\rm jet} \dot{M}_{\rm accr} c^2$. The value of $q_{\rm jet}$ is taken as a free parameter. In order to reproduce the  MAGIC observations we have that $q_{\rm jet}\sim 0.1$.
The mass accretion  rate $\dot{M}_{\rm accr}$ is strongly dependent on the relative velocity of the compact object with respect to the wind. To estimate the evolution of the accretion rate onto the black hole we have taken into account the azimuthal wind velocity $v_{\phi}\sim1.1\times10^7 (R_\star/r)/ \sin i$ cm s$^{-1}$ (Gregory \& Neish 2002, Casares et al. 2005) in addition to the radial velocity, $v_{\rm r} = 3\times 10^5 (r/R_*)^{n-2}$ cm s$^{-1}$. The Bondi-Hoyle accretion regime\footnote{We notice that the close approach of the stars during the periastron might induce a transient Roche-lobe overflow.} was considered to obtain the values of $\dot{M}_{\rm accr}$ shown in Figure 1. This is just a rough approximation used by several authors (Mart\'{\i} and Paredes 1995, Gregory \& Neish 2002). Close to the compact object there should be a transition to disk accretion, but the absence of disk signatures in the X-ray spectrum suggests that the accretion disk in LS~I+61 303 is small. We shall assume that changes in $\dot{M}_{\rm accr}$ are propagated to the jet on timescales much shorter than the orbital period.

\begin{figure}[h]
\centering 
\includegraphics{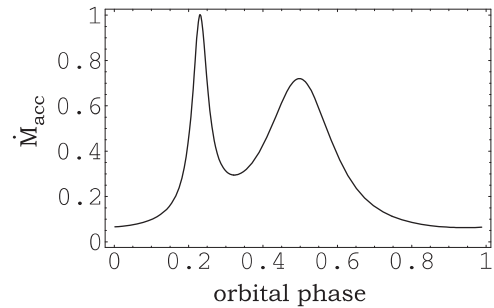}
\caption{Normalized mass accretion rate. At the periastron passage (phase 0.23) it is $\dot{M}_{\rm acc}=1.7\times10^{19}$ g s$^{-1}$.
}\label{Macc}
\end{figure}

\section{The very high-energy spectrum and light curve}
\label{sec:2}

\begin{figure}
\centering 
\includegraphics{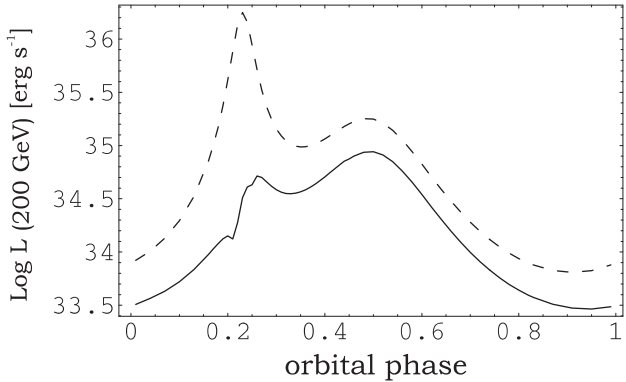}
\caption{Calculated $\gamma$-ray luminosity from hadronic interactions. The dashed line corresponds to the generated luminosity. In solid line, we show the luminosity corrected by absorption in the stellar and disk photon fields, according to the values of $\tau$ shown in Figure \ref{tau-200}. 
}\label{c-luz}
\end{figure}

In Figure \ref{c-luz} we show the gamma-ray luminosity expected from $pp\rightarrow pp+\pi^0$ and the subsequent neutral pion decays. The light curve is plotted for photon energies $\sim 200$ GeV, which is the lower energy reported by MAGIC for LS~I +61 303. The differential $\gamma$-ray emissivity is calculated applying the $\delta$-function approximation (Aharonian \& Atoyan 2000). The jet is assumed to expand in a conical way (see parameters in Table \ref{tab}). The magnetic field ($B\leq 10^8$ G at the base of the jet) results from the equipartition condition (see Bosch Ramon et al. 2006a) and then decreases with the distance as $\propto z^{-2}$. Notice that the field will change with the accretion rate and hence, with the orbital phase. Since hadronic energy losses are negligible, the maximum proton energy has been obtained at the jet formation point equating the shock acceleration rate to the proton synchrotron rate. This leads to $\gamma_p=7\times 10^6$ at the base of the jet during the periastron passage. Photohadron interactions in the stellar field (the typical energy of the stellar photons is $\sim 1$ eV) are well below the threshold even for the most energetic protons. These type of losses are possible with X-ray photons form the accretion disk, but the X-ray luminosity is low enough as to neglect them in a first approximation.  The differential distribution of relativistic protons has an index $\alpha=2.5$ in order to match the slope of the MAGIC spectrum. We note that the usual particle acceleration through the Fermi mechanism can lead to such a soft distribution for shocks with low Mach number. Actually, $\alpha$ could be time dependent, affected at some extent by the strong variations in the accretion rate.

\begin{figure}[h]
\centering 
\includegraphics[angle=-90,width=110mm]{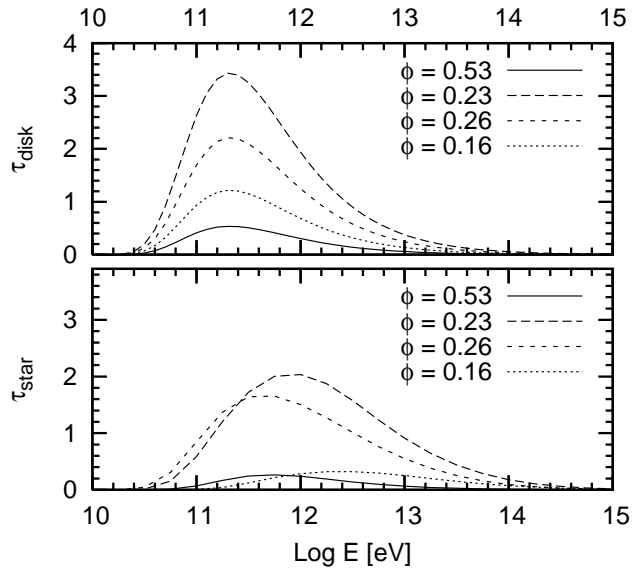}
\caption{Optical depth as function of the $\gamma$-ray photon energy. It was calculated for fixed values of the orbital phase $\phi$ indicated in the figure. The lower panel shows absorption by the stellar photon field, and the upper one, by the photon field of the circumstellar disk.
}\label{tau-E}
\end{figure}

\begin{figure}[h]
\centering 
\includegraphics{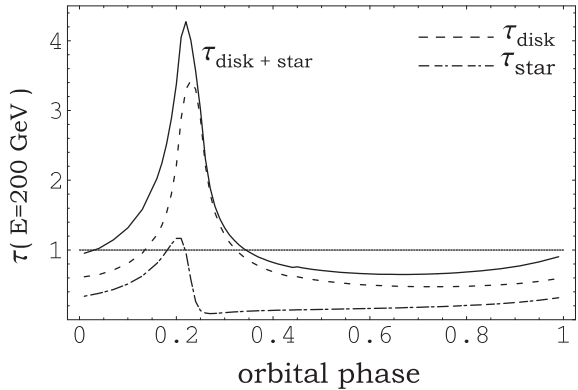}
\caption{Optical depth for the propagation of $\gamma$-ray photons with energy $E=200$ GeV.
}\label{tau-200}
\end{figure}

There are two maxima in the light curve shown in Figure \ref{c-luz}, one at the periastron and the other around phase 0.5, when the relative velocity is minimum. The solid line shows the luminosity corrected by $\gamma \gamma$ absorption in the stellar and disk radiation fields. We have calculated the opacity of the anisotropic stellar field as in Dubus (2006a), taking into account angular effects and the finite size of the star. The orientation of the orbit is given by Casares et al. (2005). The circumstellar disk is modeled as a blackbody of $T_{\rm disk}=17000$ K for the inner region (up to an inner radius of 3 stellar radii) and then with an emissivity that goes as $\propto\rho_{\rm w}^2$ (Waters 1986, Bosch-Ramon et al. 2006b). The disk farther truncates at 12 $R_\star$. The inner region of the disk is normalized to emit $L_{\rm disk}\sim 2 \times 10^{37}$ erg s$^{-1}$, which is a non negligible fraction of the total thermal emission of the system (Casares et al. 2005). The total $\gamma \gamma$ cross-section has been considered for interactions with photons from the disk. The energy dependence of the optical depth from both disk and stellar contribution is shown in Figure \ref{tau-E}. Figure \ref{tau-200} presents the variation of the total optical depth with the orbital phase for a photon of energy of 200 GeV.
 
\begin{figure}
\centering 
\includegraphics{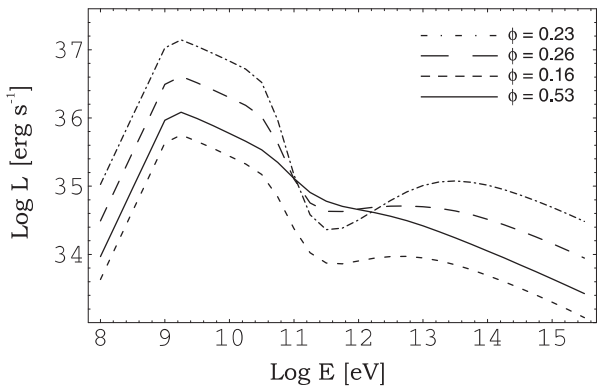}
\caption{Estimated spectral energy distribution, affected by absorption, for different values of the orbital phase. In this plot $q_{\rm jet}=1$. Note that at $\phi=0.53$ (solid line) the expected signal should be stronger than at the periastron passage ($\phi=0.23$) for the range of energies covered by MAGIC. The actual cutoff for high-energy photons is around 1 PeV.}
\label{spectros}
\end{figure}

The spectral energy distribution (SED) from $pp$ interactions, estimated for different orbital phases, is shown in Figure \ref{spectros}. Close to the periastron the absorption results to be significant, and mainly dominated by the circumstellar disk emission. On the other hand, at phase $\phi=0.53$ most of the radiation escapes from the source, making it detectable at this phase and not during the first accretion peak, when the ambient wind density is maximum. Notice that the more realistic modeling of the wind and the absorption changes these results from those previously presented by Romero et al. (2005) \footnote{Also, we considered in the present case the usual convention for the orbital phase as time proportional. In Romero et al. (2005) we have used the orbital parameter as linearly related to the so-called true anomaly.}.

We have computed the high-energy $\gamma$-ray spectra resulting from cascades traversing the anisotropic stellar radiation field during the periastron passage. The star is characterized by a surface temperature  $T_\star=26000$ K, and a stellar radius $R_\star=10$ R$_\odot$. Monte Carlo simulations were performed after developing a computational code based on the scheme outlined by
Protheroe (1986) and Protheroe et al. (1992). The $\gamma$-ray spectra produced through IC  interactions were approximated as in Jones (1968), modified in similar way to Bednarek (1997). We introduced the effects of the finite size of the star and the spatial variation of the ambient photon field density by considering the geometric configuration as in Dubus (2006a). The magnetic field in the cascade region, originated in the star, was assumed to be lower than 0.1 G, assuring in this way a strongly Compton dominated regime. 
We followed the cascades induced by the injected gamma-ray flux with mentioned photon index $\alpha=2.5$, during the periastron passage (i.e. orbital separation of $\sim 2.56\,R_\star$). 

\begin{figure}
\centering \includegraphics{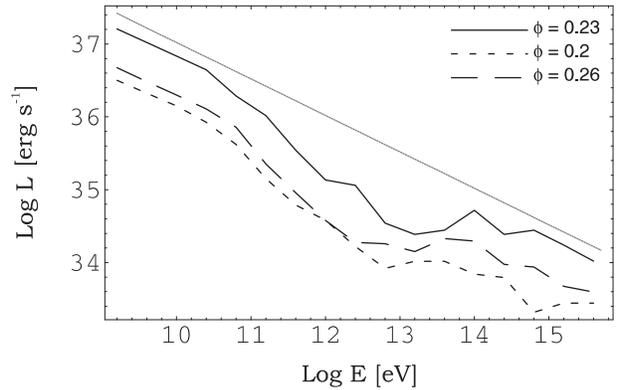}
\caption{Emerging spectral energy distribution after the IC dominated cascade developed in the stellar photon field (the effect of the decretion disk of the Be star is not included). The straight line only indicates the slope of the injected spectrum, with photon index $\alpha= 2.5$ (which mimics the original proton spectral index). These results were obtained through Monte Carlo simulations, properly taking into account the geometric configuration. The curves were normalized with $q_{\rm jet}=1$ and the actual cutoff is around 1 PeV.}\label{cascados}
\end{figure}

In Figure \ref{cascados} we present the results of the simulations. The one-dimensional cascade development is calculated in the anisotropic stellar field along the line of sight, forming a 30 deg angle with the jet axis. At phase 0.5 the opacity is not enough to sustain a local cascade (gamma-rays that escape in the direction of the star can produce cascades in the stellar photosphere, and some gamma-rays could, in principle, be redirected toward the observer, see Bednarek 2006a and 2006b). We notice that at the periastron passage the effect of the cascades is to produce a softer $\gamma$-ray spectrum with index $\alpha\sim 2.8$ in the MAGIC energy range. Such a feature should be detectable through larger exposures than those reported in Albert et al. (2006).

\section{Discussion}
\label{sec:3}
The luminosity obtained at phase 0.5 implies that the constant coupling the jet kinetic power and the accretion luminosity should be $q_{\rm jet}\sim 0.1$ in order to explain the $7\times 10^{33}$ erg s$^{-1}$ inferred from MAGIC detections. Powerful jets can be present in microquasars, as showed by Gallo et al. (2005) for the case of Cygnus X-1.

At energies below the break of the hadronic spectrum ($\sim$ GeV)
 the emission of leptonic origin should change the shape of the SED yielding the EGRET  spectrum. The hadronic interactions that produce neutral pions also generate relativistic electrons and positrons by the decay of charged $\pi$-mesons. We have not computed here the emission of such leptons, for we are concerned only with the very high energy range. Some estimations of the synchrotron luminosity of secondaries were presented in Romero et al. (2005).

Both leptonic and hadronic microquasar models for the gamma-ray emission in LS~I +61 303 can reproduce the main features of the observed gamma-ray emission. The detection of the source at phase 0.5 is a consequence of the opacity effects and is independent of the dominant radiative process in the jet. Even in hadronic models like the one presented here, near the periastron passage gamma-rays will initiate electromagnetic cascades with IC emission, hence the actual situation could be a mixture of radiative processes. More realistic simulations of the photospheric cascades should include the effects of circumstellar disk absorption. 

In a pure leptonic model the radiative losses equating the acceleration rate fix the maximum energies of the $\gamma$-ray photons at $\sim 1$ TeV (Bosch-Ramon et al. 2006b). Detection of $\gamma$-ray emission of LS~I +61 303 at higher energies than those observed by MAGIC (e.g. at $E\geq 10$ TeV) could give support to a hadronic model, and to the presence of intense magnetic fields at the base of the jet. 

A clear distinction about the nature of most of the primary gamma-rays can be established through the detection of neutrinos, which are produced only in hadronic models. In the context of our model, we expect around 3-4 muon neutrinos per squared km per year at Earth, so the source might be detectable by IceCube (Christiansen et al. 2006, see also the discussion in Torres \& Halzen 2006). 

As we mentioned in the Introduction, an alternative model for the gamma-ray production in LS~I +61 303 involves an energetic pulsar which could generate a strong wind that would stop the accretion. Particles might be accelerated at the colliding wind region, producing gamma-rays through IC and $pp$ interactions (Dubus 2006b, Chernyakova et al. 2006). However, it is not clear how the morphological and variability properties of the system might be explained in this scenario. What is clear is that the high accretion rates and the observed low X-ray luminosity seem to exclude a scenario based upon an accreting pulsar, since there is no trace of the emission from the heated surface.    
Future observations with MAGIC will help to detect the source close to the periastron and the spectral evolution along the orbit, providing in this way more constraints to the models. 



\begin{acknowledgements}
We thank Valenti Bosch-Ramon, Wlodek Bednarek, E. Derishev, and Josep M. Paredes for discussions and an anonymous referee for valuable comments. This work was supported by grants PICT 03-13291, BID 1728/OC-AR (ANPCyT) and PIP 5375 (CONICET).
\end{acknowledgements}


\end{document}